\documentclass[12pt]{iopart}
\usepackage{epsfig}

\def\S{\cal S\mathit}

\def\Hz{\rm\, Hz}

\newcommand{\be}{\begin{equation}}
\newcommand{\ee}{\end{equation}}
\newcommand{\ba}{\begin{eqnarray}}
\newcommand{\ea}{\end{eqnarray}}

\begin{document}

\title[On line power spectra
  identification and whitening]{On line power spectra
  identification and whitening for the noise in interferometric gravitational
  wave detectors}

\author{Elena Cuoco\dag\footnote[4]{To whom correspondence should be addressed (cuoco@fi.infn.it)},Giovanni Calamai\ddag, Leonardo Fabbroni\dag, Giovanni Losurdo*, Massimo Mazzoni\dag, Ruggero Stanga\dag,  Flavio Vetrano\P\ }
\address{\dag\ Dipartimento di Astronomia e Scienze dello Spazio, Universit\'a
  di Firenze and INFN Firenze/Urbino Section}

\address{\ddag\ Osservatorio Astrofisico di Arcetri and INFN Firenze/Urbino Section}

\address{*\ INFN Firenze/Urbino Section}

\address{\P\  Universit\`a di Urbino and INFN Firenze/Urbino Section}

\begin{abstract}
The knowledge of the noise Power Spectral Density of  interferometric detector of gravitational waves is fundamental for detection algorithms and for the analysis of the data.
In this paper we address both to the problem of identifying the noise Power
Spectral Density of
interferometric detectors by parametric techniques and to the problem of the
whitening procedure of the sequence of data.
We will concentrate the study on a Power Spectral Density like the one of the
Italian-French detector VIRGO and we show that with a reasonable  number of parameters we succeed in modeling a spectrum like the theoretical one of VIRGO, reproducing all its features.

We propose also the use of adaptive techniques to identify and to whiten on
line the data of interferometric detectors. We analyze the behavior of the
 adaptive techniques in the field of stochastic gradient and  in the
 Least Squares ones. 
As a result, we find that the Least Squares Lattice filter is the
best among those we have analyzed. It optimally succeeds in
following all the peaks of the noise power spectrum, 
and one of its outputs is the whitened part of the spectrum.
Besides, the fast convergence of this algorithm let us follow the slow non
stationarity of the noise.
These procedures could be used to whiten the overall power spectrum
or only some region of it. The advantage of the techniques we propose is
that they do not require  a priori knowledge of the noise power spectrum to be
analyzed. Moreover the adaptive techniques let us identify and remove the
spectral line, without building any physical model of the source that have
produced them. 

\end{abstract}


\submitto{\CQG}   


\maketitle
\section{Introduction}
\subsection{Generalities}
The detection of gravitational waves represents a major goal in contemporary Physics. A world-wide effort 
has been made in building detectors (especially ground-based long-arms detectors) with sensitivity enough to
 make astrophysical observations or, in a wider sense, to move in the field of
 gravitational astronomy~\cite{Saulson, Schutz, Grishchuk}. The
 building of these large interferometers is now reaching the phase of data taking: TAMA (Japanese)~\cite{Ando} is
 already working; GEO (British/German)~\cite{Luck} will begin to take data next year; LIGO (U.S.A.)~\cite{Coles}in 2002; VIRGO (French/Italian)~\cite{Marion} in 2003. However, long test run of the Central Interferometer in VIRGO is foreseen during 2001, leading thus to a big amount of data to be analyzed: although these data are mainly for diagnostic purposes, they provide a very good opportunity to examine analysis techniques. In the following, even though we are referring to VIRGO antenna, our considerations can be in principle applied to all ground-based interferometric gravitational detectors.
Generally speaking, all these detectors are Michelson interferometers
with suitable technical additions in order to improve sensitivity
(see~\cite{Saulson, Blair, Barone} for exhaustive and up-to-date
descriptions of physics and technology involved in building up
interferometric gravitational antennae). A gravitational wave  (GW)
displaces in different way the far mirrors in the two arms, thus
shifting the interference pattern at the beam-splitter; however, the
best models we have for GW (astrophysics) sources  are leading to a
very small value for the wave amplitude on the earth~\cite{Blair, Barone}
requiring for its detection a spectral sensitivity of about $10^{-19} {\rm m}/\sqrt {\rm Hz}$
in a band of about 1 kHz, let's say from few Hz to many hundreds of Hz. But to
reach this nominal sensitivity is only the first challenge. In fact, other not
minor challenges arise: the run of the antenna should be continuous because we
are searching for rare time-limited events  (supernovae bursts, coalescing
binary systems) or, on the contrary, we have to integrate small continuous
signals over long time (e.g. pulsars signals), while a very sensitive
monitoring of the environment and instrument noise should be carefully and
continuously done; when considering the low value of Signal to Noise Ratio
(SNR), we are leading to foresee that large amount of data will be handled,
very big computing power (TFLOPS) will be requested, large archiving capacity
(TBYTES) will be prepared for storage and retrieval~\cite{Owen}. Finally, from 
a preliminary point of view for data treatment, which is indeed the argument
of this paper, major challenge relies on the fact that the noise of GW
detectors does not satisfy the simplifying assumptions of white noise, but
expected  noise is a colored broad-band background with some spectral
(deterministic) peaks. Moreover the noise distribution could be non
stationary and non Gaussian. 
\subsection{The problem}
\begin{figure}
\begin{center}
\epsfig{file=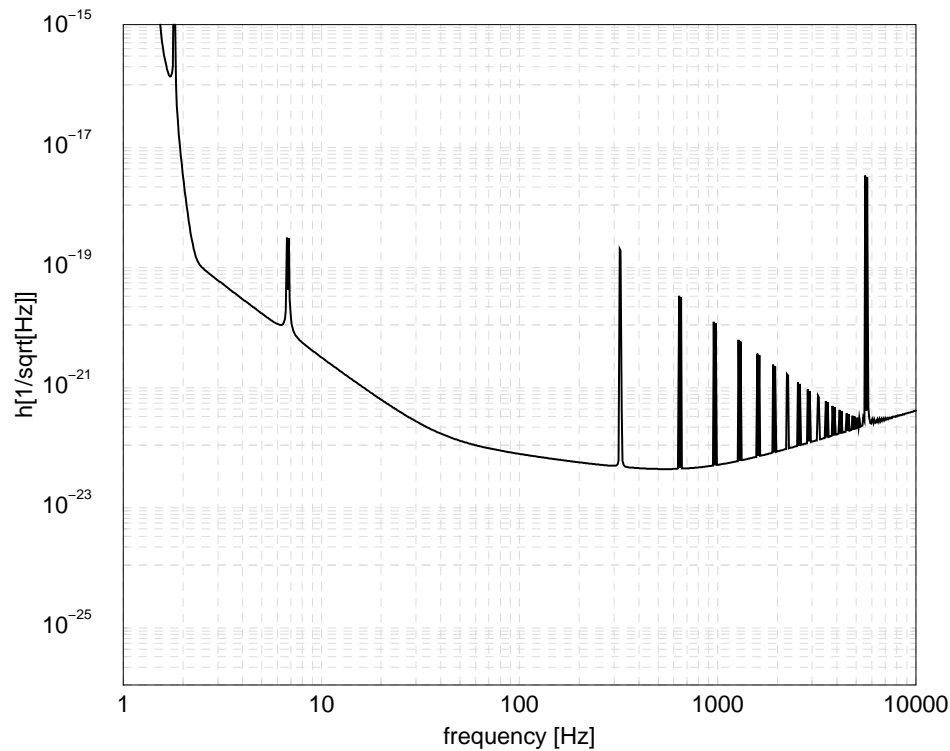,angle=270,width=0.8\textwidth}
\end{center}
\caption{VIRGO sensitivity curve}
 \label{fig:VIRGO}
\end{figure}

The large amount of data produced by gravitational wave detectors will be
essentially noise  and, hopefully, buried in noise there will be the signal we
are looking for. As we already saw, ground-based interferometric detectors are sensible to a
broad  band frequencies (2-3Hz to more than 1kHz) in revealing relative displacement of test masses at
the near  and far extremities  of interferometer arms due (possibly!) to GW signal, but unfortunately a lot of other factors can cause
a similar displacement. The test masses are suspended to pendular
structures in order to isolate them from  seismic noise~\cite{Bradaschia}, but thermal noise of the suspension chain will cause a displacement of the
mass~\cite{Cagnoli}. Also  shot noise and radiation pressure of the laser will move the mirrors~\cite{Saulson, Meers}.
The physicists are working in  modeling  all  possible causes of noise
in the interferometer giving out a sensitivity curve of the
apparatus~\cite{Thorne1,Thorne2, Cella-Cuoco, Beccaria, Cella, Cagnoli2, Braginsky}.
This curve is limited  at very low frequencies (few Hz and below) by seismic noise; in the
middle band by  thermal noise and at high frequencies (higher than $0.7-1$kHz) by shot-noise.
In figure~\ref{fig:VIRGO} is reported the predicted VIRGO sensitivity curve obtained as incoherent sum of  all estimated noise contributions~\cite{Punturo, Punturo2}. This curve is characterized by a
broad-band noise plus several very narrow peaks due to the violin modes of the
suspensions wires, that will make the detection of a gravitational signal in
this frequency band  very difficult. For this reason efforts have been
made in the preparation of the analysis of data for cutting
~\cite{Finn, Sintes, Chassande-Mottin} out these resonances. 
It is evident that the analysis of data to detect the gravitational signal
requires an accurate knowledge of the noise, which means a statistical
characterization of the stochastic process, evaluating its stationarity and its
Gaussian nature, and  in the case of local stationarity and Gaussian nature  an
accurate estimation of the Power Spectral Density (PSD).

The output of the interferometer will be surely non stationary over a long
period of time, so we must be ready in following the changes in the PSD. A way to achieve this is to estimate the PSD on a chunk of
data at different interval of time, using classical techniques~\cite{Kay,Hayes,Therrien}. We propose to use adaptive methods to
follow on line the change in the feature of the spectrum in such a way to have
at any desired instant the correct curve for the PSD.

If we are able in identifying the noise of our detector we can also apply the
procedure of whitening of the data.

The goal of a whitening procedure is to make the sequence of data
delta-correlated, removing all the correlation of  the noise.
 Most of the theory of detection is in the frame of a
wide-sense stationary  Gaussian white noise, but in our problem the noise is
surely a colored one and, in principle, there could be present non
stationary and non Gaussian features.
If we whiten the data, supposing hence to be in the frame of a stationary and
Gaussian noise, we can apply the optimal algorithm detection\cite{Zubakov}.

For example, if we assume that the noise data which we are analyzing are stationary and Gaussian
distributed and we suppose to know the waveform of our signal, then the optimal detection filter is the matching between our sequence
of data and the Wiener filter:

\begin{equation}
\label{eq:wienerM}
M(t,\theta )=\int _{-\infty }^{\infty }e^{-2i\pi \nu t}\frac{h(\nu ,\theta )}{S(\nu )}d\nu\, ,
\end{equation}
 where $ S(\nu ) $ is the noise PSD  and $ h(\nu ,\theta ) $
is the template of the signal we are looking for, $\theta$ the parameters of
the waveform. As it is evident from (\ref{eq:wienerM}) the operation of whitening is implicitly done each time we
apply the Wiener filter to detect a signal, because we weight the data with
the inverse of PSD of noise: in such a way we have a 'whitened' sequence to
analyze.

Moreover when we are searching a transient signal of unknown form it is very
important to  have a whitened noise~\cite{Hello, Pradier}. 
The importance of whitening data is also linked to the possibility of reducing
their dynamical range~\cite{Allen, GRASP}.

The aim of this paper is to show how to identify the noise PSD and how to whiten the data produced by an interferometric
detector before applying any algorithm detection.

In the section~\ref{sec:parametric} we underline the advantages in parametric modeling; in
section~\ref{sec:whitening} we show the whitening techniques based on a
lattice structure. In section~\ref{sec:Virgoresults} we report the application
of PSD fitting and data whitening on VIRGO-like simulated data.
In section~\ref{sec:adaptive} we introduce the theory of adaptive filters
based on stochastic and least squares methods, and its application to
VIRGO-like data: for this we compare their performances on simulated data.

\section{Parametric Modeling}
\label{sec:parametric}
The advantages of parametric modeling with respect to the classical spectral
methods are described in an exhaustive way in reference \cite{Kay}. 
We want here to underline that  the kind of analysis we have to perform can
take advantage from these methods for two main reason.
First of all we can achieve a better resolution in the estimation of PSD,
because we can use in a better way the information of the autocorrelation
function. In fact we suppose that the process we are analyzing is governed by a
dynamical law and we can use the knowledge of autocorrelation function until a certain lag
and then extrapolate its value under the dynamical hypothesis we made.
Moreover we may compress the information of the PSD in a
restricted number of parameter and not in the full autocorrelation function.
This can help, for example, if we want to create a data base of noise sources.

In this context we want to talk about parametric modeling also because it offers the possibility to write down a linear whitening filter and to build in a fast way simulated
data on which we  perform our tests of whitening filter.

We work in the field of rational functions to fit the theoretical PSD. We will show that it is possible to obtain a fit of
the theoretical PSD of an interferometer output like the one of VIRGO   with an autoregressive
moving average (ARMA) or autoregressive (AR) model\cite{cuoco, cuoco2, cuoco3}, then we will
use the data we can generate in this way, to test the whitening algorithms we propose.
\subsection{ARMA and AR models}
The procedure to estimate the PSD using parametric modeling is based on three
steps: 

\begin{enumerate}
\item select the appropriate model for the process; 
\item estimate the model parameters from the given data; 
\item use these parameters in the theoretical power spectrum density for the model.
\end{enumerate}
Once we have the parameters which make the fit we use them to generate noise
data to perform our tests.

A general process described by a ARMA(P,Q) model, being $P$ the number of poles and $Q$ the number of zeros, satisfies the relation: 
\begin{equation}
\label{eq:arma}
x[n]=-\sum _{k=1}^{P}a[k]x[n-k]+\sum _{k=0}^{Q}b[k]w[n-k]
\end{equation}
 and its transfer function is given by 
\begin{equation}
{\mathcal{H}}(z)=\frac{{\mathcal{B}}(z)}{{\mathcal{A}}(z)}\, ,
\end{equation}
 where$ {\mathcal{A}}(z)=\sum _{k=0}^{P}a[k]z^{-k} $ and $ {\mathcal{B}}(z)=\sum _{k=0}^{Q}b[k]z^{-k} $
.

The PSD of the ARMA output process is 
\begin{equation}
P_{ARMA}(f)=\sigma ^{2}|\frac{B(f)}{A(f)}|^{2}\, ,
\end{equation}
$ \sigma  $ being the variance of driven white noise $ w $, $ A(f)={\mathcal{A}}(2\pi if) $
and $ B(f)={\mathcal{B}}(2\pi if) $.
An autoregressive process AR(P) is governed by the relation
\begin{equation}
\label{eq:ar}
x[n]=-\sum _{k=1}^{P}a[k]x[n-k]+w[n]\, ,
\end{equation}
and its PSD for a process of order $ P $ is given by
\begin{equation}
P_{AR}(f)=\frac{\sigma ^{2}}{|1+\sum _{k=1}^{P}a_{k}\exp (-i2\pi kf)|^{2}}
\end{equation}
Once we selected the model for our process, we need to find the parameters
for this model. The parameters of the ARMA model are linked to the autocorrelation
function of the process by the Yule-Walker equations\cite{Kay}. In the general
case of an ARMA process we must solve a set of non linear equations while, if
we specialize to an AR process, that is an all-poles model, the equations to
be solved to find the AR parameters become linear.

 The relationship between the parameters of the AR model and the autocorrelation
function $ r_{xx}(n) $ is given by the Yule--Walker equations 
\begin{equation}
r_{xx}[k]=\left\{ \begin{array}{ll}
-\sum _{l=1}^{P}a_{l}r_{xx}[k-l] & \mbox {for}\, \, k\geq 1\\
-\sum _{l=1}^{P}a_{l}r_{xx}[-l]+\sigma ^{2} & \mbox {for}\, \, k=0\, .
\end{array}\right. 
\end{equation}

The problem of determining the AR parameters is the same of that of finding
the optimal ``weights vector'' ${\mathbf w}=w_k$, for $k=1,...P$  for the
problem of linear prediction~\cite{Kay}. In the
linear prediction we would predict the sample $ x[n] $ using the $ P $
previous observed data ${\mathbf x}[n]=\{x[n-1],x[n-2]\ldots x[n-P]\} $
building the estimate as a transversal filter:
\begin{equation}
\hat{x}[n]=\sum _{k=1}^{P}w_{k}x[n-k]\, .
\end{equation}

We choose the coefficients of the linear predictor by minimizing  a cost
function that is the mean squares
error $ \epsilon ={\mathcal{E}}[e[n]^{2}] $ (${\mathcal{E}}$ is the expectation operator ), being 
\begin{equation}
\label{eq:error}
e[n]=x[n]-\hat{x}[n]
\end{equation}
the error we make in this prediction and obtaining the so called Normal or
Wiener-Hopf  equations
\begin{equation}
\label{eq:Normal}
\epsilon _{min}=r_{xx}[0]-\sum _{k=1}^{P}w_{k}r_{xx}[-k]\, ,
\end{equation}
 which are identical to the Yule--Walker equations with 
\begin{eqnarray}
w_{k} & = &-a_{k}\\
\epsilon _{min} & = & \sigma ^{2}
\end{eqnarray}

This relationship between AR model and linear prediction assures us to obtain
a filter which is stable and causal~\cite{Kay}. It is this relation between
AR process and linear predictor that becomes important in the building of whitening
filter.

\subsection{\label{subsec:durbin}Durbin algorithm and lattice structure }

A method of solving the Yule--Walker equation is the Durbin algorithm~\cite{Alex}. Let us suppose the process being an autoregressive one of order $P$.  

The strategy of this method is to assume that the optimal $ (P-1) $th order filter has previously been computed, and then to calculate the optimal $ P $th
order filter based on this assumption. To accomplish the algorithm we perform a loop on the order of the process for $ 1\leq j\leq P $.
We initialize the mean squares error as $ \epsilon _{0}=r_{xx}[0] $ and then we begin the iteration on the loop, introducing the reflection coefficients $ k_{p} $~\cite{Therrien} at the stage $p$ \footnote{Note that $p$ indicates any order of the filter we choose for our model. We can always add a new stage to the filter, having an AR($p+1$) model}, which are linked to the autocorrelation function  $r_{xx}$ and to the $a_j$ parameters of the filter at the stage $p-1$ by the relation :
\begin{equation}
k_{p}=\frac{1}{\epsilon _{p-1}}\left[ r_{xx}[p]-\sum _{j=1}^{p-1}a^{(p-1)}_{j}r_{xx}[p-j]\right] \, .
\end{equation}
So the loop for $ 1\leq j\leq P $ proceeds in the following way: 

\begin{itemize}
\item estimation of the reflection coefficient
\begin{equation}
k_{j}=\frac{1}{\epsilon _{j-1}}\left[ r_{xx}[j]-\sum _{i=1}^{j-1}a^{(j-1)}_{i}r_{xx}[j-i]\right] \, .
\end{equation}
\item at the $ j $th stage the AR parameter of the model is equal to the $ j $th reflection coefficient
\begin{equation}
a_{j}^{(j)}=k_{j} \, ,
\end{equation}

\item the other parameters are updated in the following way:
\end{itemize}
For $ 1\leq i\leq j-1 $
\begin{equation}
a_{i}^{(j)}=a_{i}^{(j-1)}-k_{j}a_{j-i}^{(j-1)}
\end{equation}
\begin{equation}
\epsilon _{j}=(1-k_{j}^{2})\epsilon _{j-1}
\end{equation}

\begin{itemize}
\item At the end of the $ j $ loop, when $ j=P $, the final AR parameters are
\begin{equation}
a_{j}=a_{j}^{(P)},\qquad \sigma ^{2}=\epsilon _{P}\, .
\end{equation}

\end{itemize}

\section{Whitening Filter}
\label{sec:whitening}
\subsection{Link between AR model and whitening filter}

\begin{figure}
\begin{center}
\epsfig{file=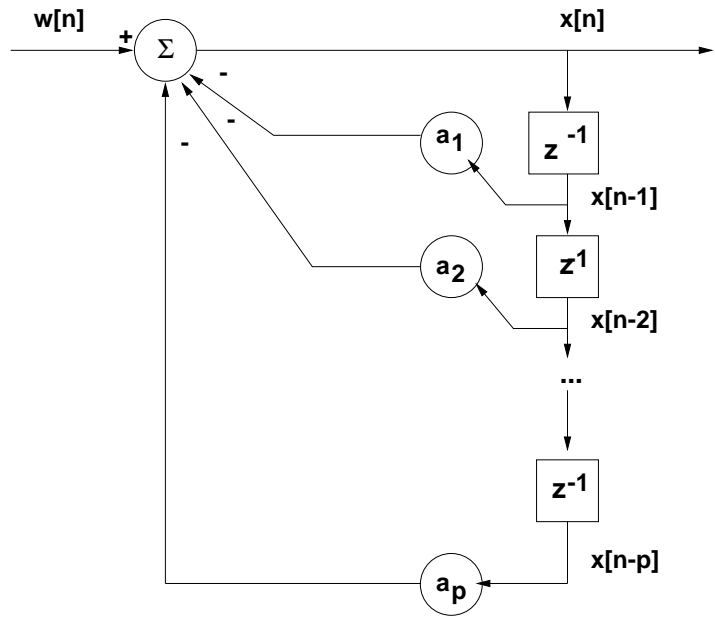,width=0.6\textwidth}
\end{center}
\caption{Whitening filter and AR filter.}
\label{fig:ar-predic}
\end{figure}

The tight relation between the AR filter and the whitening filter is clear in
the figure~\ref{fig:ar-predic}. The figure describes how an AR process colors
a white process at the input of the filter if you look at the picture from left
to right. If you read the picture from right to left you see a colored process
at the input that pass through the AR inverse filter coming out as a white process.

When we find the $ P $ parameters that fit a PSD of a noise process, what
we are doing is to find the optimal vector of weights that let us reproduce the process
at the time $ n $ knowing the process at the $ P $ previous time. All
the methods that involves this estimation try to make the error signal (see
equation (\ref{eq:error}) ) a white process in such a way to throw out all the correlation
between the data (which we use for the estimation of the parameters).

The Durbin algorithm  introduces in a natural way the Lattice structure for the whitening
filter.

We show how the reflection coefficients  $k_p$ are used to build a lattice
whitening filter.
Let us suppose to have a stochastic Gaussian and stationary process  
$x[n]$ which we modeled as an autoregressive process of order $P$.
We define the {\it forward} error (FPE) for the filter of order $P$ in the
following way 
\begin{equation}
e_P^f[n]=x[n]+\sum_{k=1}^P a_k^{(P)} x[n-k]\, ,
\end{equation}
where the coefficients $a_k$ are the coefficients for the AR model for the
process $x[n]$. The FPE represents the output of our filter.
We can write the $zeta$ transform for the FPE at each stage
$p$ for the filter of order  $P$ as
\begin{equation}
  \label{eq:FPEzeta}
  FPE(z)=F_p^f[z]X[z]=\left(1+\sum_{j=1}^{p}a_j^{(p)}z^{-j}\right)X[z]\, .
\end{equation}
At each stage $p$ of the Durbin algorithm the 
coefficients $a_p$ are updated as
\begin{equation}
\label{eq:k2a}
a_j^{(p)}=a_j^{(p-1)}+k_p a_{p-j}^{(p-1)}\qquad 1\le j\le p-1 \ .
\end{equation}
If we use the above relation for the transform $F_p^f[z]$, we obtain
\begin{equation}
\label{eq:FPEzeta_rec}
  F_p^f[z]= F_{p-1}^f[z]+k_p\left[z^{-p}+\sum_{j=1}^{p-1}a_{p-j}^{(p-1)}z^{-j}\right]\ .
\end{equation}
Now we introduce in a natural way the 
{\it backward} error of prediction BPE
\begin{equation}
  \label{eq:BPEzeta_rec}
  F_{p-1}^b[z]= z^{-(p-1)}+\sum_{j=1}^{p-1}a_{p-j}^{(p-1)}z^{-(j-1)}\ .
\end{equation}
In order to understand the meaning of $F_p^b[z]$ let us see its action in the time domain
\begin{eqnarray}
  \label{eq:BPEtime}
F_{p-1}^b[z]x[n]=e_{p-1}^b[n]=\\ \nonumber
x[n-p+1]+\sum_{j=1}^{p-1}a_{p-j}^{(p-1)}x[n-j+1]\, .
\end{eqnarray}
So $e_{p-1}^b[n]$ is the error we make, in a backward way, in the prediction
of the data $x[n-p+1]$ using  $p-1$ successive data
$\{x[n],x[n-1],\ldots, x[n-p+2]\}$.
 
We can write the eq.~(\ref{eq:FPEzeta_rec}) using $F_{p-1}^b[z]$.  
Let us substitute this relation in the z-transform of the filter $F_p^f[z]$
\begin{equation}
\label{eq:FPEzeta_R}
   F_p^f[z]= F_{p-1}^f[z]+k_pF^b_{p-1}[z].
\end{equation} 
In order to know the FPE filter at the stage  $p$ we must know  the  BPE
filter at  the stage $p-1$. 

\begin{figure}
\begin{center}
\epsfig{file=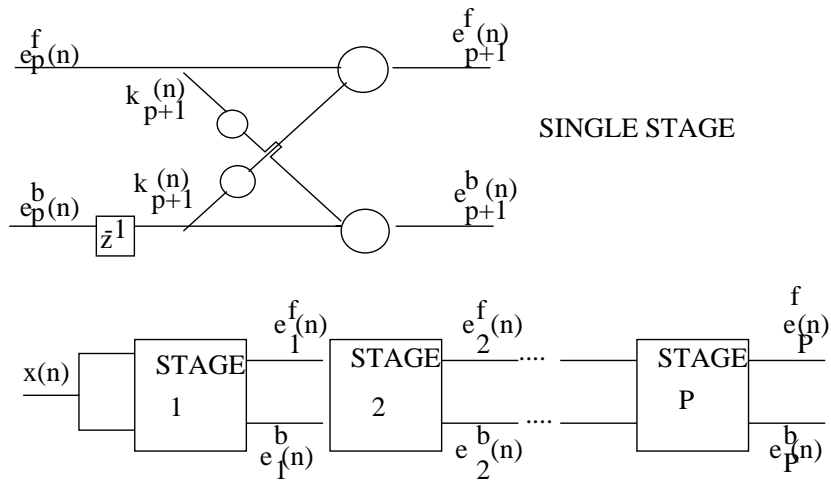,angle=270,width=0.7\textwidth}
\end{center}
\caption{Lattice structure for Durbin filter.}
\label{fig:lattice}
\end{figure}

Also for the {\it backward} error we may write in a similar way the relation 
\begin{equation}
  \label{eq:BPEzeta_R}
  F_p^b[z]=  z^{-1}F_{p-1}^b[z]+k_p F_{p-1}^f[z]\ .
\end{equation}
The equations~(\ref{eq:FPEzeta_R})~(\ref{eq:BPEzeta_R}) represent our lattice
filter that in the time domain could be written 
\begin{eqnarray}
e_p^f[n]&=& e_{p-1}^f[n]+k_p e_{p-1}^b[n-1]\, ,\\
e_p^b[n]&=& e_{p-1}^b[n-1]+k_p e_{p-1}^f[n]\, .
\end{eqnarray}
In figure~\ref{fig:lattice} is showed how the lattice structure is used to
estimate the forward and backward errors.

Using a lattice structure we can implement the whitening filter following
these steps: 
\begin{itemize}
\item estimate the values of the autocorrelation function  $\hat
r_{xx}[k], \mbox{$ 0\le k\le P$}$ of our process $x[n]$;
\item use  the Durbin algorithm to find the reflection coefficients $k_p, \mbox{$ 1\le p\le P$}$;
\item implementation of the lattice filter with these coefficients $k_p$
  initiating the filter  $e_0^f[n]=e_0^b[n]=x[n]$.
\end{itemize}
In this way the forward error at the stage  $P$-th is equivalent to the
forward error of a transversal filter and represents the output of the
whitening filter.

The procedure of whitening will be accomplished before applying the algorithms
for the detection of gravitational signal of different wave forms. The level
of whiteness of the data needed for the various algorithms could be different.
It is important to have a common language and to assign a parameter which characterizes
the performance of whitening filter. We want now to introduce this parameter
that let us quantify the level of whiteness of data at the output of whitening
filter.

\subsection{\label{sec:whiteness}The ``whiteness'' of data: measure of flatness of PSD}

The spectral flatness measure for a PSD is defined as\cite{Kay} 

\begin{equation}
\label{eq:flatness}
\xi =\frac{\exp (\frac{1}{N_{s}}\int _{-N_{s}/2}^{N_{s}/2}\ln (P(f))df)}{\frac{1}{N_{s}}\int _{-N_{s}/2}^{N_{s}/2}P(f)df}
\end{equation}

where the integral is extended in the bandwidth of Nyquist frequency; this parameter
satisfies

\begin{equation}
\label{eq:limits_flatness}
0\leq \xi \leq 1
\end{equation}

If $ P(f) $ is very peaky, then $\xi\simeq 0$, if $ P(f) $ is flat than
$\xi=1$. 

With th definition~\ref {eq:flatness} the flatness for a process at the output
of a whitening filter built with a minimum phase filter (as the AR filter is)
is

\begin{equation}
\label{eq:xie}
\xi _{e}=\xi \frac{r_{xx}[0]}{r_{ee}[0]},
\end{equation}
 where $ r_{xx}[0] $ and $ r_{ee}[0] $ are the values of the autocorrelation
function of the process before and after the whitening procedure, and $ \xi  $ is the value of flatness for the initial sequence~\cite{Kay}.

\section{Results on simulated VIRGO-like noise data}
\label{sec:Virgoresults}
We want to investigate the performance of the Durbin filter in fitting the
VIRGO PSD and in whitening the simulated output of this interferometer.

We can simulate the data as AR or ARMA process~\cite{cuoco}, by fitting the
theoretical PSD as an  AR or ARMA model. If we simulate the data as an AR
process and fit them with an AR model, the number of parameters we must
use will be small. In the real situation the output of interferometer will
not be an AR process. 
This does not mean that we cannot ever fit the data as an AR process, but that
probably we need a greater number of parameters. In order to be closest to
the real situation we use an ARMA fit to the theoretical PSD and we performed the tests following the steps:

\begin{itemize}
\item \emph{ARMA fit to theoretical VIRGO-like noise PSD}
\item \emph{Generation of noise data with the ARMA parameters}
\item \emph{realization of one process of noise}
\item \emph{P order selection for the AR fit to the realization of the noise}
\item \emph{Durbin(P) whitening filter. }
\end{itemize}
\subsection{The VIRGO noise power spectrum}
We consider a theoretical curve for a VIRGO-like power spectrum in which
shot noise and thermal noise of pendulum, 
mirrors and violin modes are present:

\be
S(f) = \frac{\displaystyle S_1}{\displaystyle f^5} + 
\frac{\displaystyle S_2}{\displaystyle f} + S_3\left( 1 + \left(\frac{f}{f_K}\right)^2\right) + S_v(f) 
\ee
where

\ba
f_K &= & 500 {\rm Hz} \qquad \mbox{shot noise cut frequency} \\
S_1 &= & 1.08\cdot 10^{-36} \qquad\mbox{pendulum mode} \\
S_2 &= & 0.33 \cdot 10^{-42} \qquad\mbox{mirror mode} \\
S_3 &= & 3.24\cdot 10^{-46} \qquad\mbox{shot noise}
\ea

The contribute of violin resonances is given by 
\be
S_v(f) = \sum_n\frac{1}{n^4}\frac{f_1^{(c)}}{f} \frac{C_c \phi_n^2}
{\left(\displaystyle
\frac{1}{n^2}\frac{f^2}{f_1^{(c)2}} -1 
\right)^2+\phi_n^2} + (c\leftrightarrow f)
\ee
where we take into account  the different masses of close and far mirrors,
being
\be
f_n^{(c)} = n \cdot 327{\Hz} \qquad
f_n^{(f)} = n \cdot 308.6\Hz 
\ee
\be
C_c  = 3.22\cdot 10^{-40}\qquad  C_f = 2.82\cdot 10^{-40} \qquad
\phi_n^2 = 10^{-7}
\ee
The difference between far and close masses leads to the presence of
double violin peaks, as we can see in figure~\ref{fig:armafit}, where
we have  plotted the spectrum obtained with a sampling
frequency $f_s=4096\Hz$.

We suppose to explore the band of frequencies from $ 10 $Hz to $ \sim 2000 $
Hz, where it is most probable to find a gravitational signal, choosing a sampling
frequency of $ 4096 $Hz. The low frequency part of the spectrum has been
filtered to cut the tail of the thermal noise. 
We used a second order high pass filter with spectral density~\cite{Orfanidis}:

\begin{equation}
\label{eq:prefiltro}
|H(\omega)|^2=\frac{1}{1+\epsilon^2_{pass}\left(\frac{\displaystyle\cot(\omega/2)}
{\displaystyle\cot(\omega_{pass}/2)}\right)^{2N}} \,
\end{equation}
with the following values
\begin{equation}
\epsilon_{pass}=1000\, ,\qquad N=2\, ,\qquad \omega_{pass}=3\pi\, .
\end{equation}

\subsection{Data simulation}

First of all we make an ARMA fit to the theoretical PSD with the techniques
used in \cite{cuoco}. We choose to use $ P=32 $ and $ Q=32 $ parameters
then we simulated the data in the time-domain using the
relation~\ref{eq:arma}, with a pre-heating techniques as described in \cite{Kay}.
\begin{figure}
\begin{center}
\epsfig{file=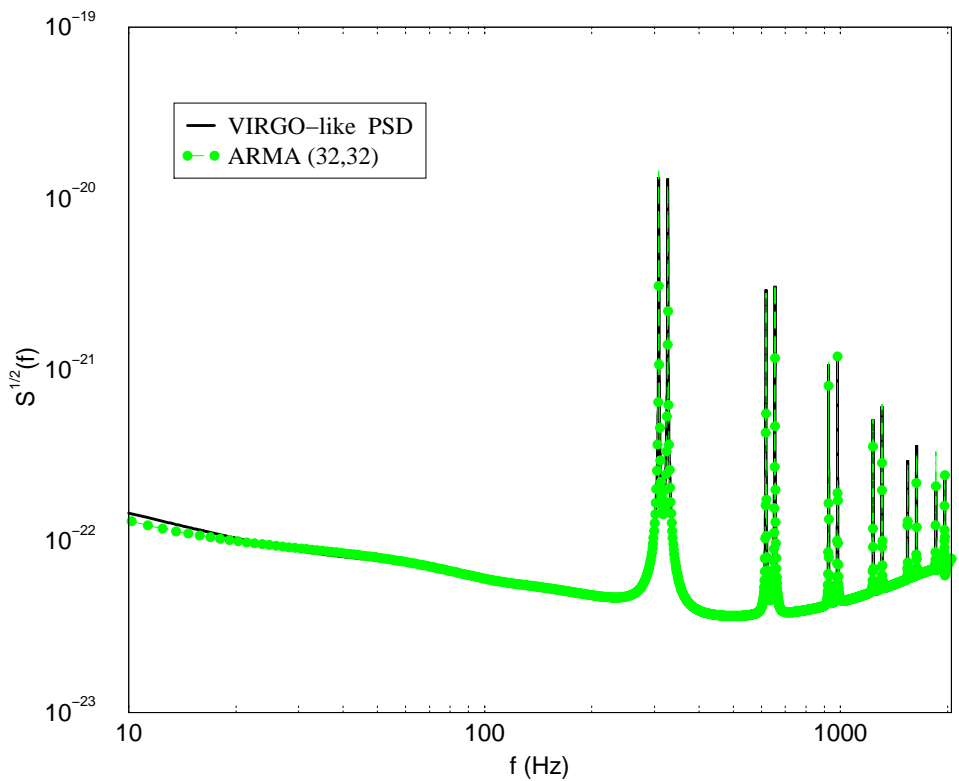,angle=270,width=0.8\textwidth}
\end{center}
\caption{\label{fig:armafit}ARMA(32,32) fit to theoretical VIRGO PSD}
\end{figure} 

\begin{figure}
\begin{center}
\epsfig{file=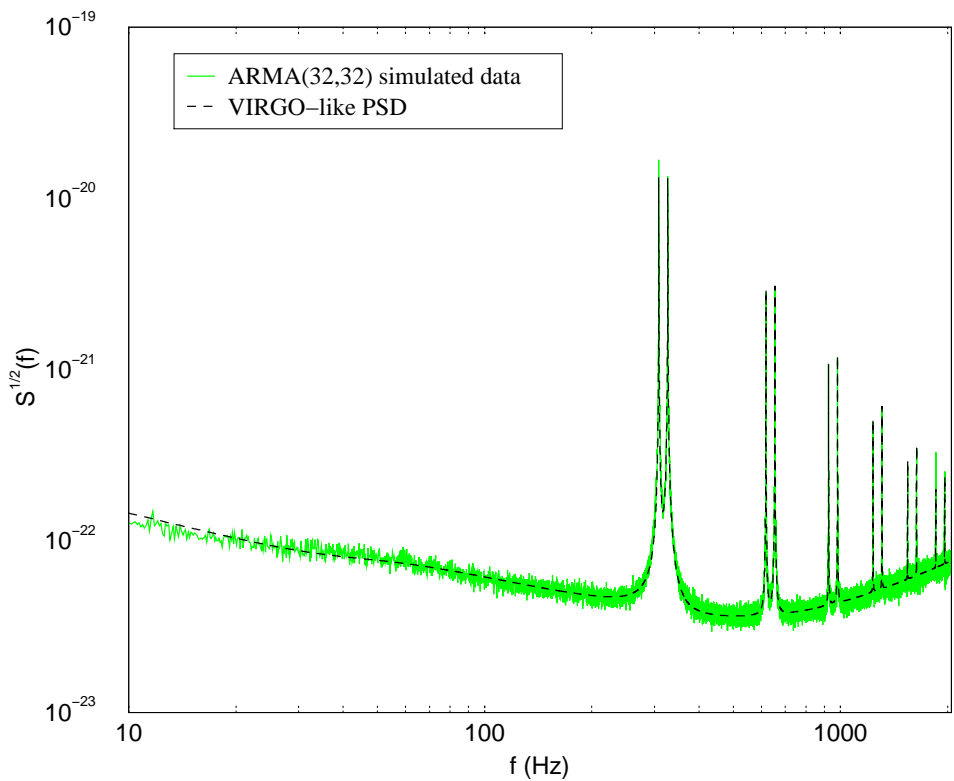,angle=270,width=0.8\textwidth}
\end{center}
\caption{\label{fig:datasim}PSD of VIRGO-like simulated data.}
\end{figure}

In figure \ref{fig:armafit} we plot the theoretical VIRGO PSD and the ARMA(32,32)
fit, while in figure \ref{fig:datasim} we show the PSD obtained as an averaged
periodogram on $ 50 $ realizations of the process for simulated data. As
it is evident the fit is good and we can suppose the time-domain data well represent
the expected Gaussian and stationary noise process for VIRGO interferometer.

\subsection{Order Selection }
The idea of the whitening filter is that the process we analyze is an
autoregressive one and that once we have the AR parameters we can use them in
the filter of figure~\ref{fig:ar-predic}.

In general we don't know the order of our process, even if we suppose that it
is an AR one.
If it is an AR of order P, and we use an order $p<P$, the fitted spectrum will
be smoother than the original one; if we  choose an order  $p>P$, there
may be spurious peaks in the spectrum.
In both cases the whitening will be not good. 

If our process is not AR, the number of parameters could be in principle
infinite. We must then fix a criterion that let us select the right order of
the process, or at least the best one.

We used the classical order selection criteria~\cite{Kay, Hayes}, that is the Akaike information
criterion (AIC), the forward prediction error (FPE) the Parzen's criterion
(CAT) and the minimum description length (MDL) one
\begin{eqnarray}
AIC(P) & = & N\log \epsilon (P)+2P\, \, ,\\
 &  & \nonumber \\
FPE(P) & = & \epsilon (P)\frac{N+P+1}{N-P-1}\, \, ,\\
 &  & \nonumber \\
CAT(P) &= &\left( \frac{1}{N}\sum _{j=1}^{P}\frac{N-j}{N\epsilon
 _{j}^{}}\right) -\frac{N-P}{N\epsilon _{P}^{}}\, \, , \\
MDL(P) & = & N\log \epsilon ^{}(P)+P\log N\, \, ,
\end{eqnarray}
where $ \epsilon (P) $ is the mean square error at the order $ P $ and
$ N $ is the length of data. 
In literature the MDL criterion is considered the best among them, because it
is robust with respect to the length of the sequence, while the others depend a
lot on N\cite{Hayes}.
Suppose, as in the real situation, we have not access to the theoretical PSD of our
noise process and we want to estimate the best order of the whitening filter.
We can use a single realization of process, i. e. a sequence of N-data to estimate the autocorrelation function
and apply the selection order criteria to it. 
The results of these criteria are reported in table~\ref{tab:virgo}.

\begin{table}
\caption{Minimum of order selection criteria on a single sequence for $1$ minute of ARMA simulated VIRGO data.\label{tab:virgo}}

\begin{indented}
\lineup
\item[]\begin{tabular}{@{}llll}
\br                           
MDL &AIC &FPE &CAT \\
\mr
338 &626 &626 &681 \\
\br
\end{tabular}
\end{indented}
\end{table}

The best order to whiten the data is given by the MDL criterium~\cite{Kay}\cite{Hayes}
which produces an order of $ 338 $ parameters. 

This number is an indicative one. We can choose to build a higher order filter
to be sure to have whitened data at the output of the filter.  We choose to adopt the number of parameters of MDL criterion
and we test the flatness of the spectrum at the output of the Durbin filter measuring the value of $ \xi . $ 

All the order selection criteria give an estimation of the
number of parameters such that  the output of the
whitening filter has  the maximum value of $ \xi  $.

\begin{figure}
\begin{center}
\epsfig{file=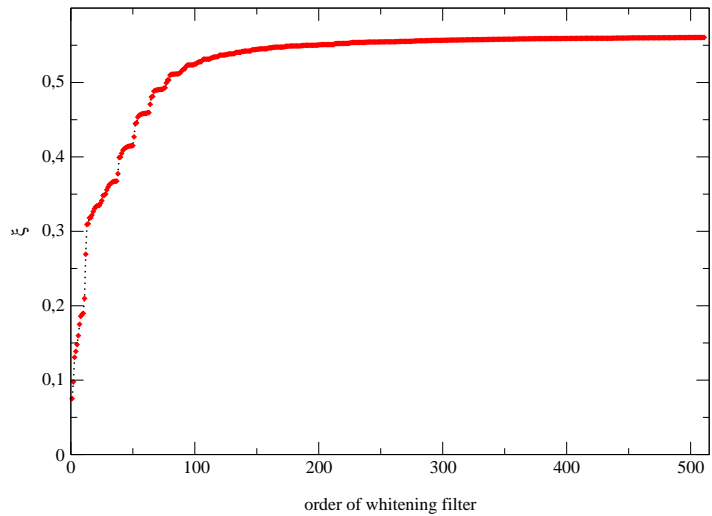,angle=270,width=0.7\textwidth}
\end{center}
\caption{\label{fig:xi(p)}Behavior of \protect$ \xi \protect $ with respect to the
order P for the VIRGO-like simulated data.}
\end{figure}

In figure \ref{fig:xi(p)} we report 
$ \xi  $ versus  the number of parameters of whitening filter. The
value of $ \xi  $ for very low order $ P $ is small and, as expected,
it increases with the order $ P $ until it converges to a plateau around $ P\sim 300 $.
So we deduce that our choice of $ P=338 $ is a good estimation of whitening filter order. 

\subsection{Results of Durbin whitening filter on simulated VIRGO-like noise data }

\begin{figure}
\begin{center}
\epsfig{file=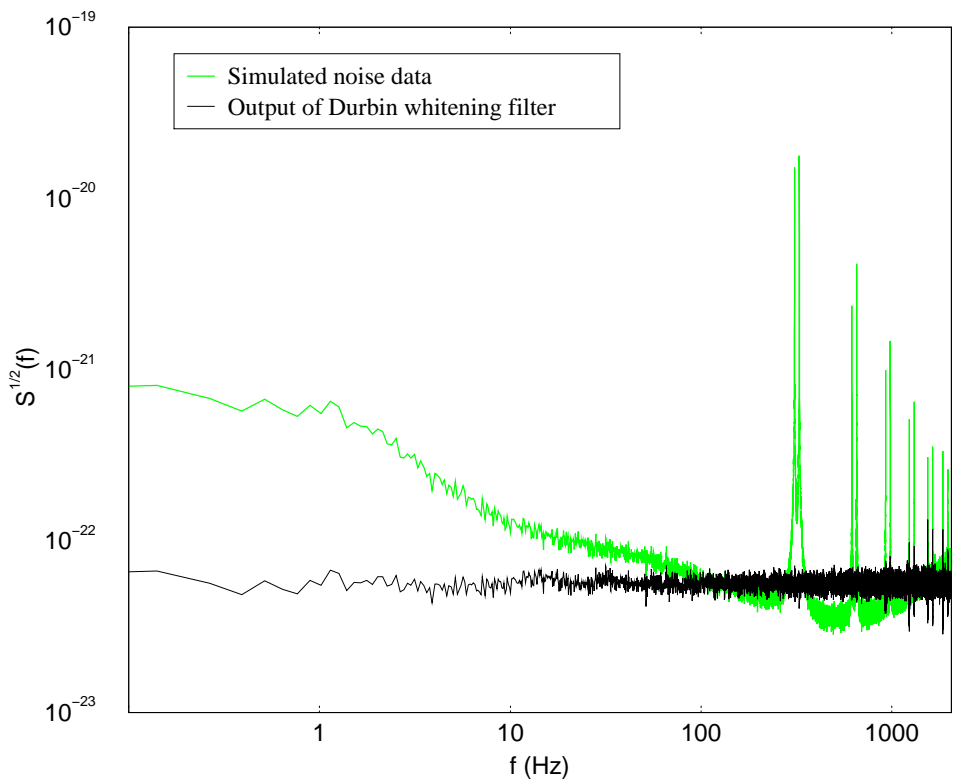,angle=270,width=0.8\textwidth}
\end{center}
\caption{\label{fig:w-durbin}Exit of Durbin whitening filter.}
\end{figure}
In figure \ref{fig:w-durbin} we plotted the averaged PSD on $100$ realizations 
of input noise and of the output of Durbin whitening filter. The results are good even if there are some residual lines at
the high frequencies. 

We can do a better whitening if we take a higher order whitening filter, but
we must pay a higher computational cost itself, because it is
proportional to the order P of the filter. The level of whitening
we choose to perform depends on the requests for the detection algorithms.

\begin{figure}
\begin{center}
\epsfig{file=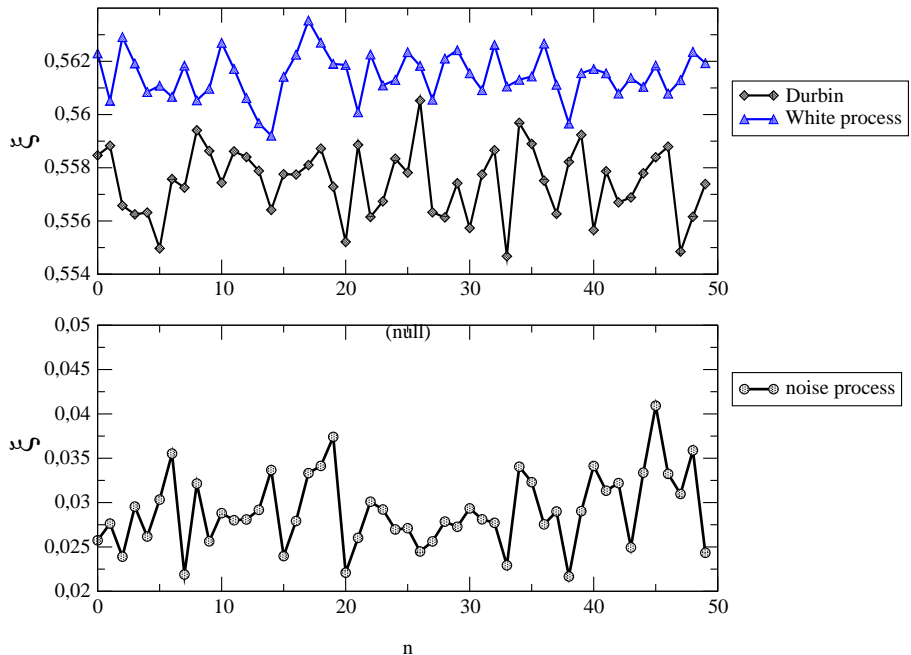,angle=270,width=0.8\textwidth}
\end{center}
\caption{\label{fig:xi}Measure of flatness for simulated noise process and outputs
of whitening filters.}
\end{figure}
 In figure \ref{fig:xi} we report the measure of $ \xi  $
in a set of realizations of the process of simulated noises. The values are reported
before any application of whitening filter and after the application of 'static'
whitening filter. We estimate the value of flatness
also for a simulated white noise process to check the goodness of the whitening
filters.

As we can see in figure \ref{fig:xi}, in a single realization the value of
flatness is high but not equal to $ 1 $, because of variance of the
estimate of periodogram. On the other hand in the averaged periodogram  the value
of $ \xi  $ is very close to $ 1 $ as we can read in table \ref{tab:xiI}. 

\begin{table}
\caption{Flatness on averaged PSD at the input and the outputs
of whitening filter for VIRGO-like simulated data.\label{tab:xiI}}
\lineup
\begin{indented}

\item[]\begin{tabular}{@{}lll}
\br
 Simulated noise&
Durbin&
White process\\
\mr
0.050&
 0.983&
 0.989\\
\br 
\end{tabular}
\end{indented}
\end{table}

\section{Adaptive filters}
\label{sec:adaptive}
\begin{figure}
\begin{center}
\epsfig{file=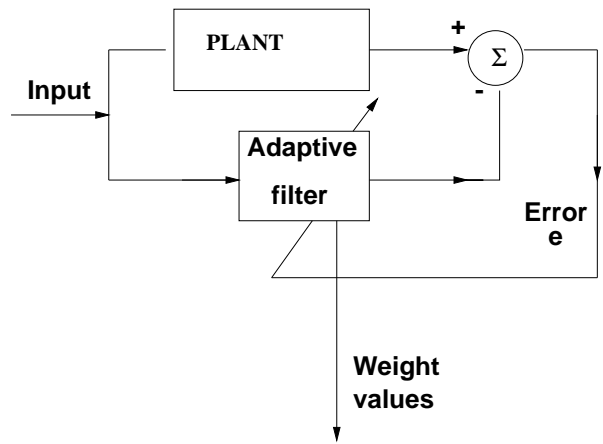,width=0.5\textwidth}
\end{center}
\caption{\label{fig:adapt}Scheme of adaptive filter for system identification.}
\end{figure}

Since now we showed filters which estimate the parameters to be used in the
whitening filter from the autocorrelation function or PSD,
i.e. these filters use  a priori information about the statistics of the
data to be analyzed. Now we want to investigate the behavior of filters which
are self-designing~\cite{Haykin}. These filters estimate the parameters directly
from data adjusting them by using as feed-back the signal obtained by the minimization
of a cost function of the error signal. In figure \ref{fig:adapt} is reported
the scheme of an adaptive filter for system identification. In our case the
plant is represented by the parameters which fit the PSD of data.
The implementation of an adaptive filter follows two steps: the filtering of
the input data and the adjustment of the filter parameters with which we process
the data to the next iteration.

The filters parameters are updated by minimizing a cost function. The way in
which we build this cost function distinguishes the adaptive methods
as~\cite{Haykin, Alex}:
\begin{itemize}
\item methods of stochastic gradient,
\item least squares methods.
\end{itemize}
To the first class belong the algorithms whose cost function is the mean
square error ${\cal E}[e^2[n]]$, where $e[n]$ is the difference between the
function we desire to find and the output of our filter.
We talk of stochastic methods because the cost function is a statistical
measure of the error.
In the second class the cost function is the weighted sum of the square errors
$e^2[n]$.  
These methods could be implemented with a block estimation or with a recursive
one (Recursive Least Squares).  
For the block estimation a block of data is acquired and then  the
least square algorithm is applied, while in the recursive one the least squares methods
should be implemented in a recursive way.

In order to be able to obtain on line the fit to the PSD, we  will
use only the recursive kind.
We used the  Gradient Adaptive Lattice (GAL), the Recursive Least Squares
(RLS) and the Least Squares Lattice (LSL). The technical details of these algorithms are described in references ~\cite{Haykin}~\cite{Alex}.

The next sections of this paper are organized in two parts: in the first part
we make a comparison of the GAL, RLS and LSL methods to fit a VIRGO-like noise
PSD. For this we simulated the data as an autoregressive
process; we  consider the parameters we use in the simulation as the true
values and we check the capability of the algorithm in converging towards these values. In the second part, we will report the application of the LSL
methods as whitening filter on data simulated as ARMA process, to show that
even in the case of a process which is not an autoregressive one, we succeed
in fitting it with an AR with a low number of parameter and in obtaining a
whitened PSD.

To check the performances of these algorithms we use the following scheme:

\begin{itemize}
\item modeling of the VIRGO as an AR process by Durbin algorithm;

\item data simulation of the noise process following the AR  relation
\begin{equation}
   x[n]=-\sum_{k=1}^P a_kx[n-k]+ w[n]\, ;
\end{equation}
\item implementation of adaptive algorithm without any information on the
  input sequence of data;
\item comparison between the estimated PSD and the obtained one.
\end{itemize}
It is fundamental to test the time convergence of the algorithms to compare
them with the typical times of non stationarities.
To this goal we measure the number of iterations by which the measured values reach the true values of the parameters. This is done in
the case of AR simulation, where the two quantities are comparable.

If we simulate the process as an AR one the MDL order selection criterion
gives as best order the value $292$. We select this one to perform our
simulation and our tests.

\section{Application to VIRGO-like simulated data}
We applied the GAL method on simulated data to verify its capability in
identifying the VIRGO-like noise power spectrum.

\begin{figure}
\begin{center}
   \epsfig{file=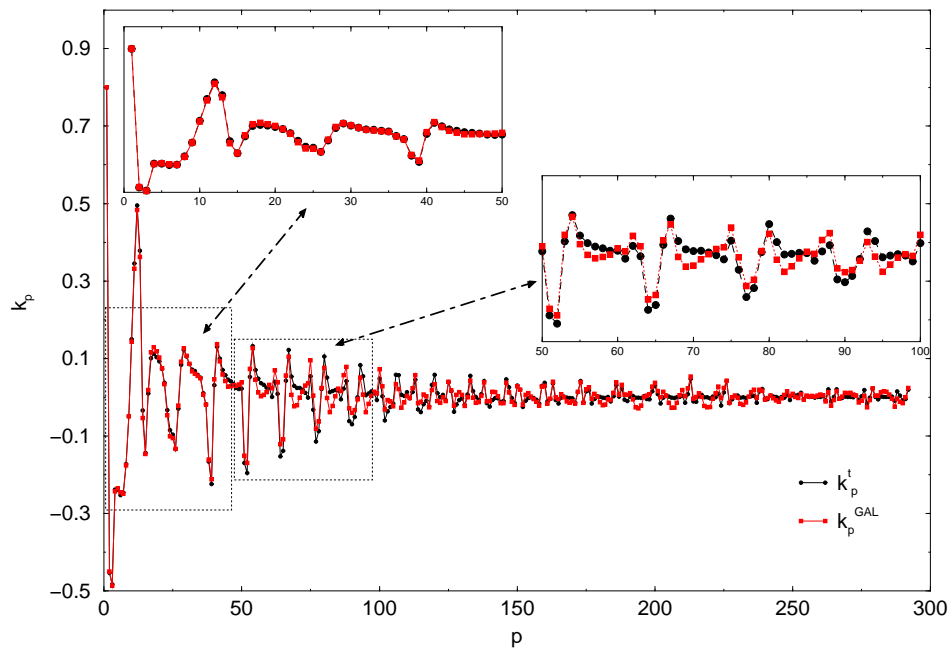,angle=270,width=0.8\textwidth}
\end{center}
 \caption{Comparison between the reflection coefficients estimated by Durbin
   and GAL algorithms}
    \label{fig:pargal}
\end{figure}

\begin{figure}
\begin{center}
 \epsfig{file=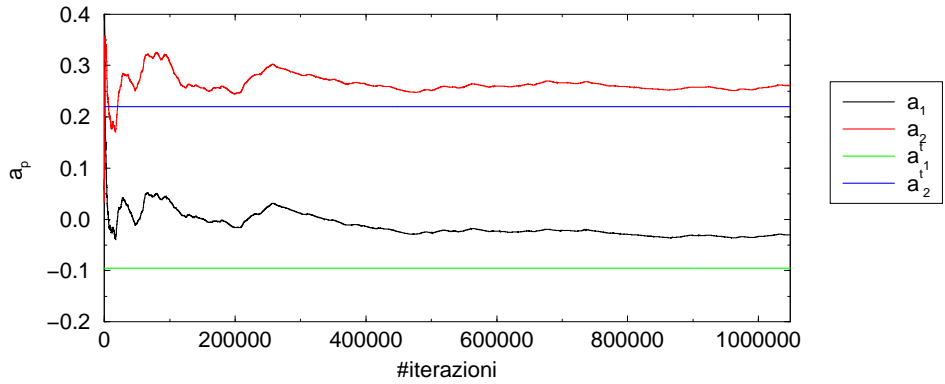,angle=270,width=0.8\textwidth}
\end{center}
 \caption{Convergence of the AR coefficients to true values after
   $4$ minutes of data}
    \label{fig:argal}
\end{figure}

\begin{figure}
   \epsfig{file=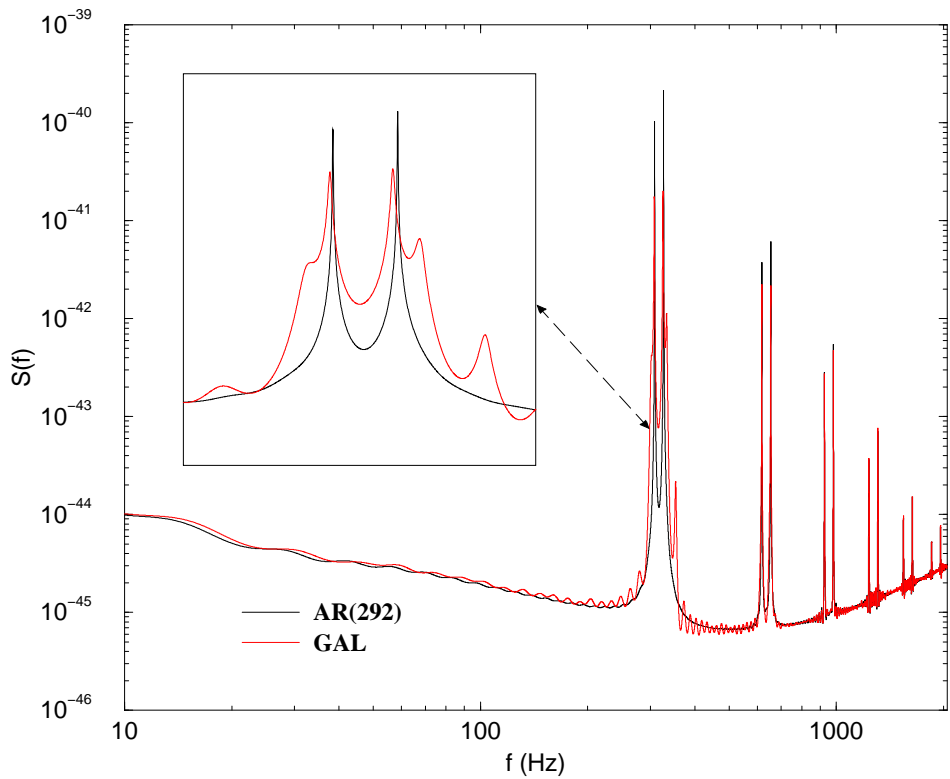,angle=270,width=0.8\textwidth}
 \caption{GAL fit to the VIRGO PSD}
    \label{fig:gal}
\end{figure}

The convergence is reached after $2$ minutes of data, but not for all the coefficients, as it is
evident in figure~\ref{fig:pargal}, where we plotted all the $292$
coefficients and zoomed the regions corresponding to $p=1,2\ldots,50$ e
$p=50,51,\ldots,100$. After the first $50$ points there is an evident discrepancy
between the simulated and the estimated reflection coefficients. This causes
the non convergence of the AR parameters even for the first two coefficients
(see figure~\ref{fig:argal}).
Even if we use a larger number of iteration, the convergence is not reached.
This is reflected in the estimation of the PSD, as you can see in the
figure~\ref{fig:gal} where the estimated GAL PSD is reported. The violin peaks
are reproduced only in a rough way. This is due to the kind of cost function
we used to find the reflection coefficient, that is optimal only in a
statistical sense and not for the actual value of the error function.

\section{\label{subsec:lsl}Least Square based methods}
The Least Squares based methods build their cost function using all the information
contained in the error function at each step, writing it as the sum
of the error at each step up to the iteration $n$ : 
\begin{equation}
\label{rls1}
\epsilon [n]=\sum _{i=1}^{n}\lambda ^{n-i}e^{2}(i|n)\, ,
\end{equation}
 
being 
\begin{equation}
\label{rls2}
e(i|n)=d[i]-\sum _{k=1}^{N}x_{i-k}w_{k}[n],
\end{equation}

where $ d $ is the  signal to be estimated, $ x $ are the data
of the process and $ w $ the weights of the filter.
We introduced the forgetting factor $ \lambda  $ that let us tune the
learning rate of the algorithm. This coefficient can help when there are non
stationary data in the data and we want that the algorithm have a short memory.
If we have stationary data we fix $ \lambda =1 $. 

There are two ways to implement the Least Squares methods
for the spectral estimation: in a recursive way (Recursive Least Squares or
Kalman Filters) or in a Lattice Filters using fast techniques~\cite{Alex}. The first kind of algorithm, examined in~\cite{cuoco2}, has a computational cost
proportional to the square of the order of filter, while the cost of the
second one is linear in the order $ P. $ 

\subsection{RLS: application to VIRGO noise data}

We used about $1$ minute of data, choosing a sampling frequency of $4096\Hz$.  

\begin{figure}
   \epsfig{file=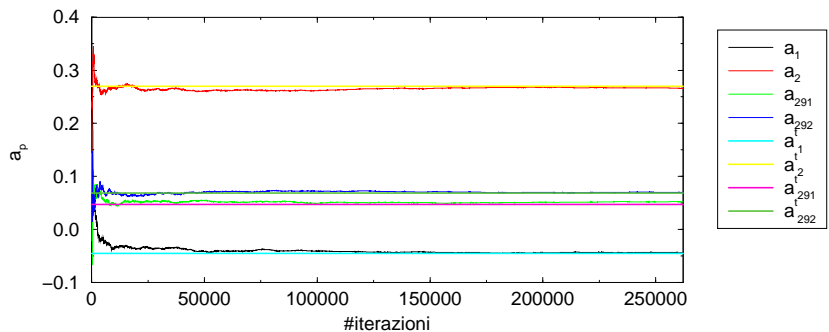,angle=270,width=0.7\textwidth}
 \caption{Convergence of the first two and the last two AR parameters for RLS filter. }
    \label{fig:ar_rls}
\end{figure}

In this transversal filter we update directly the weights, without estimating
the reflection coefficients.
In figure~\ref{fig:ar_rls} we report the convergence curves for the first two 
coefficients $a_1$, $a_2$ and for the last two coefficients 
$a_{291}$, $a_{292}$ estimated by the RLS algorithms and the corresponding 'true' values $a^t_1$, $a^t_2$, $a^t_{291}$ and $a^t_{292}$.
\begin{figure}
   \epsfig{file=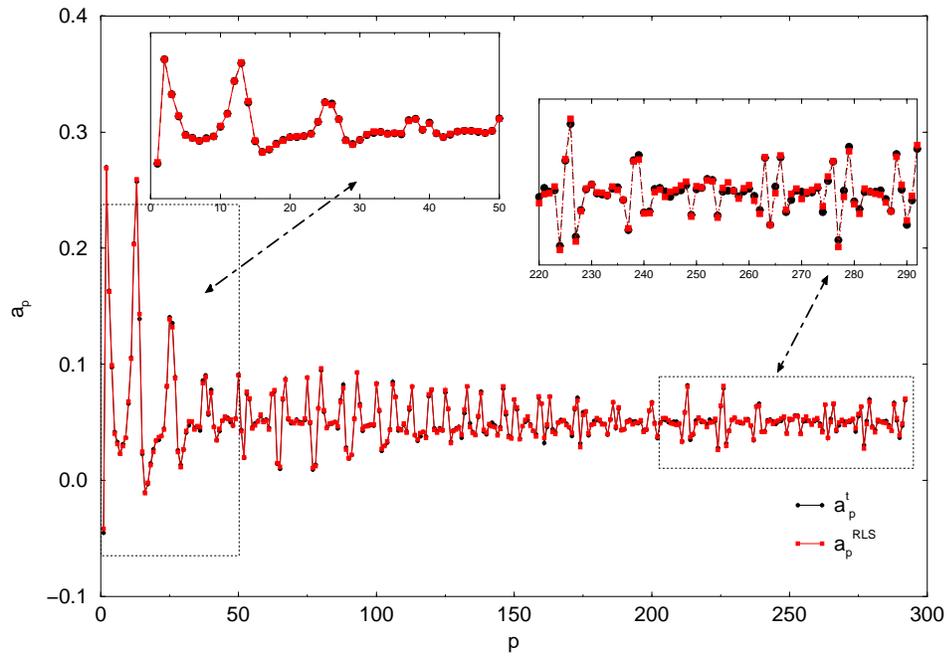,angle=270,width=0.8\textwidth}
 \caption{ Simulated AR parameters and estimated coefficients by  RLS algorithm  }
    \label{fig:parrls}
\end{figure}
The RLS algorithm converges to the true value of the parameters, and its
convergence time its of the order of $30$ sec.

\begin{figure}
   \epsfig{file=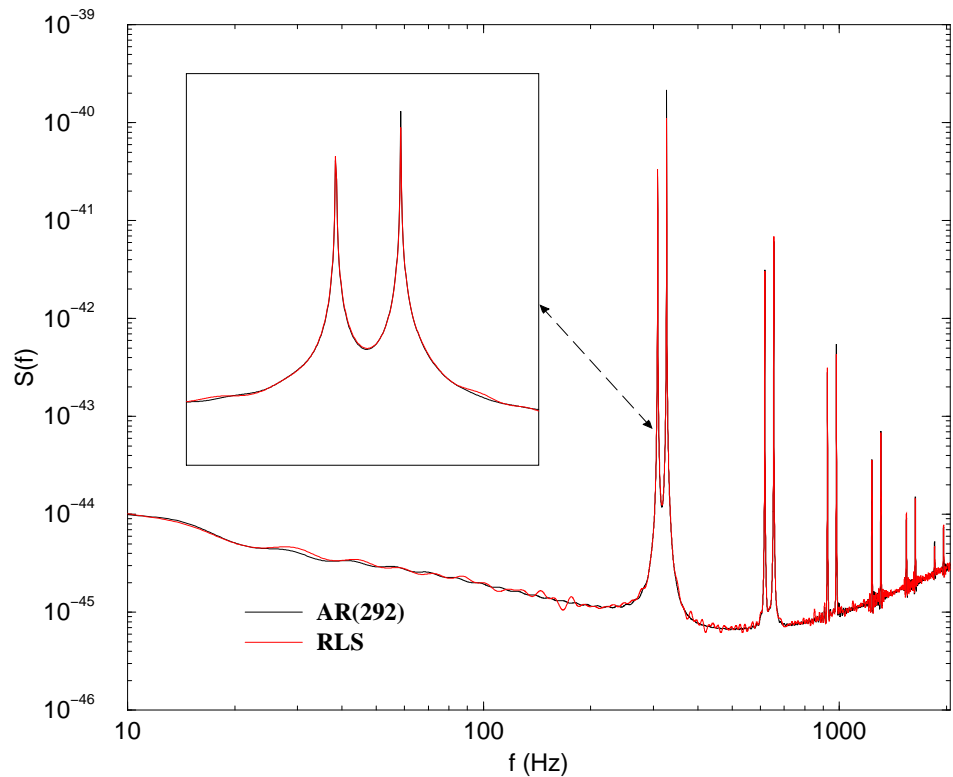,angle=270,width=0.8\textwidth}
 \caption{RLS fit to the VIRGO PSD}
    \label{fig:rls}
\end{figure}

In figure~\ref{fig:parrls} we report the $292$ parameters for the AR model 
estimated with RLS after $1$ minute of iterations and the corresponding true
values. In the zooms we can see the first $50$ and
the last $70$ coefficients. There is a small discrepancy in the estimations of
the last coefficients, but this doesn't affect the fit of the original PSD
as it is evident in figure~\ref{fig:rls}, where all the spectral features are
well reproduced.

\subsection{LSL: application to VIRGO-like noise data}
The computational cost of RLS is prohibitive for an on line implementation.
Moreover its structure is not modular as for the GAL algorithm, thus forcing
the choose of the order $P$ once for all.
The algorithm with a modular structure like that of the lattice  offers the
advantages of giving an output of the filter at each stage $p$,  so  in
principle we can change the order of the filter by imposing some criteria on
its output. On the contrary the Least Square Lattice filter is a modular
filter with a computational cost proportional to the order $P$.

We introduced for the LSL filter the forward and backward reflection
coefficients. These in principle could have different values if the sequence
is not stationary, but we simulate the VIRGO-like noise data as a stationary
process, therefore  $k_p^f=k_p^b=k_p$. Moreover we use the pre-windowed case $\lambda=1$.

We used always $1$ minute of data with a sampling frequency of  $4096\Hz$.
The AR parameters have been estimated from the reflection coefficients using
the relation~(\ref{eq:k2a}).

The error $e_p^f$ at the last stage is the whitened sequence of the input data.
So at the output of LSL filter we find the parameter for the estimation of the
AR fit to VIRGO PSD and the whitened sequence of data.

\begin{figure}
\epsfig{file=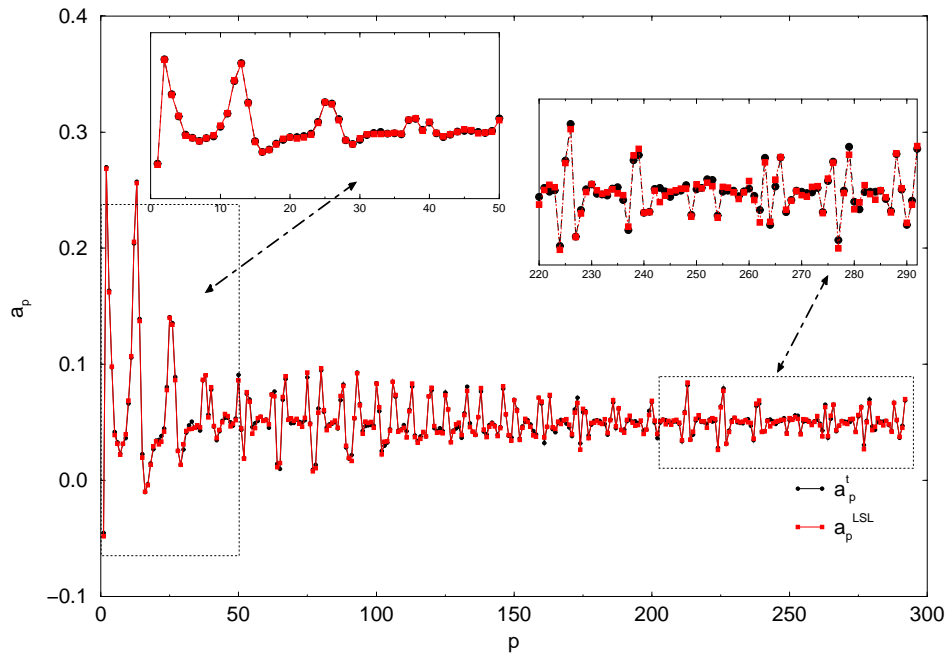,angle=270,width=0.8\textwidth}
\caption{LSL estimated AR parameters.\label{fig:parlsl}}
\end{figure}

As we expected, the performances are similar to the ones of RLS filter and the
convergence is reached after about $30\sec$ of data. This behavior is satisfied by all the coefficients  $a_p$ as it is evident in
figure~\ref{fig:parlsl} where we plotted all the coefficients of simulation
and the estimated ones.

\begin{figure}
\epsfig{file=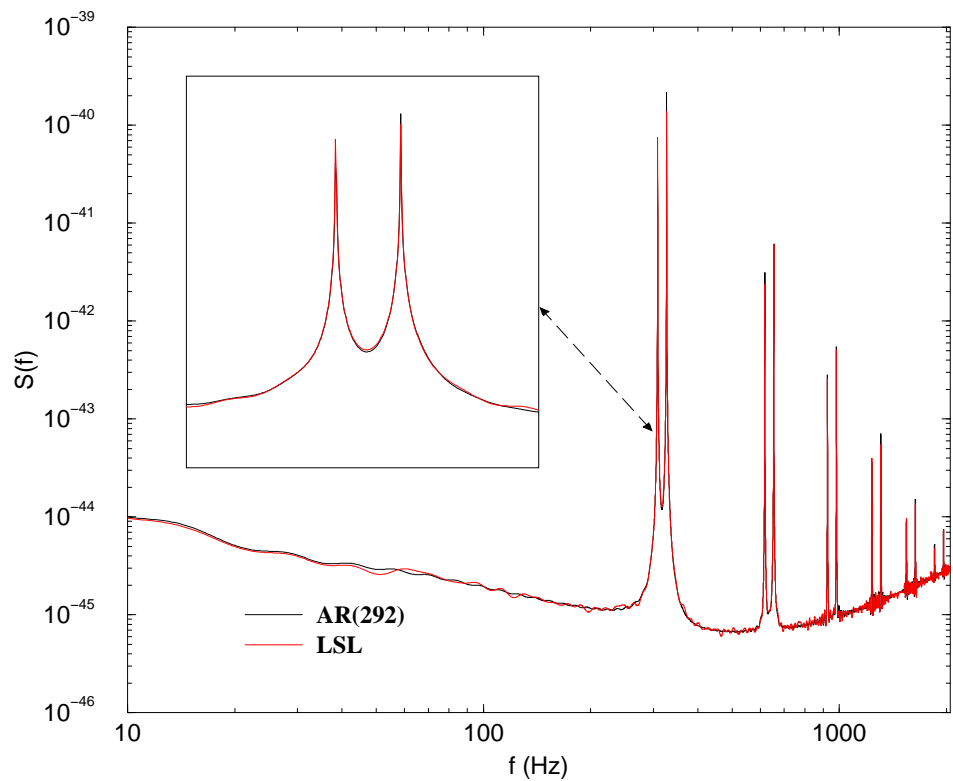,angle=270,width=0.8\textwidth}
\caption{\label{fig:lslspettro}LSL fit to the VIRGO-like noise PSD.}
\end{figure}
We zoomed on to the first $50$
coefficients and to the last $70$ ones. As for the RLS filter there is a
small discrepancy only for the last coefficients, but this does not affect the
spectral estimation as reported in figure~\ref{fig:lslspettro}, where all the
violin peaks are well reproduced.
In only one minute of data we succeeded in identifying an AR model with  $292$
parameters. 
If we think about not stationarity noise with characteristic time of  one hour, we are sure to obtain on line the right estimation of the PSD, and the right
whitened sequence.

\begin{figure}
\epsfig{file=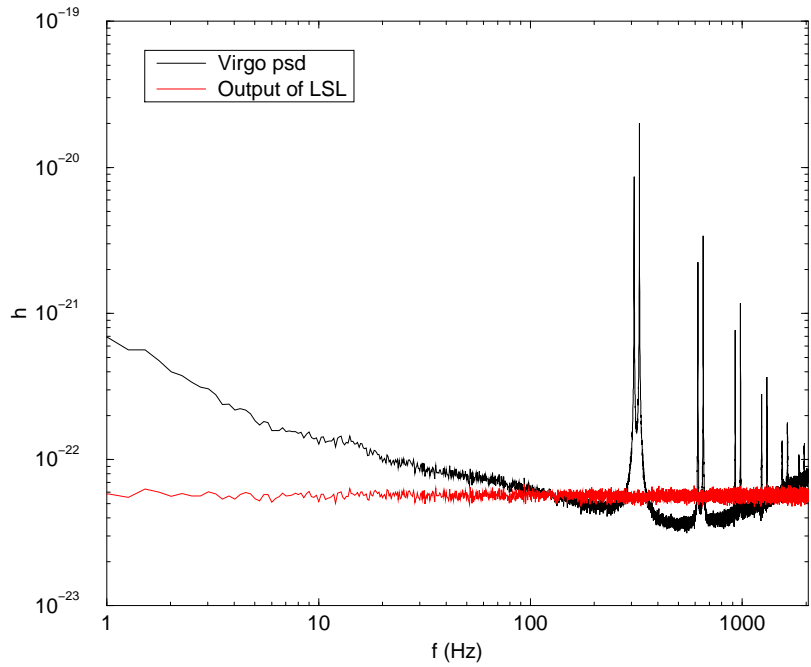,angle=270,width=0.8\textwidth}
\caption{\label{fig:lsl_old}Exit of LSL whitening filter.}
\end{figure}

In figure~\ref{fig:lsl_old} we reported  the PSD of the sequence 
$e_P^f[n]$ obtained as averaged periodogram on  $100$ noise simulations.

It is clear that the LSL is a good whitening filter and it offers the
advantages, with respect to the Durbin one, of being adaptive and of working without
estimating before the autocorrelation function from the data.

\subsection{LSL statistics}
In order to evaluate the goodness of an estimator we verify if it is an
unbiased one and if it satisfies the Cramer-Rao bound which for the AR
parameters is~\cite{Kay}
\begin{equation}
\label{eq:Cra}
var(\hat{a}_i) \ge \frac{\sigma^2}{N} [{\bf R}_{xx}^{-1}]_{ii}\qquad
\mbox{$i=1,2,\ldots p$} \, ,
\end{equation}
and
\begin{equation}
var(\hat{\sigma}) \ge \frac{2\sigma^2}{N}\, .
\end{equation}
\begin{figure}
\epsfig{file=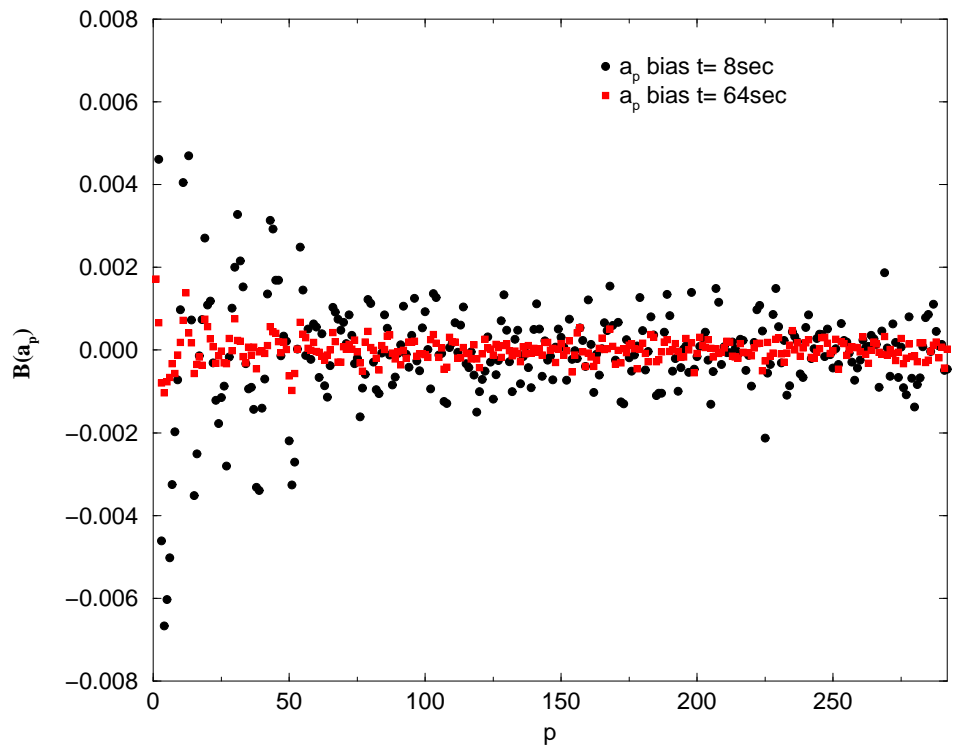,angle=270,width=0.8\textwidth}
\caption{\label{fig:biasa}Bias for the AR parameters.}
\end{figure}
We fixed $1$ minute of data as the maximum length for $N$  and we estimated the
bias for the parameters $a_p$. The statistical quantities have been evaluated
as averages on $100$ realizations of the process. In figure~\ref{fig:biasa}
we report the bias for each AR parameter 
\begin{equation}
  \label{eq:biasa}
  B(a_p)={\cal E}[\hat a_p]-a^t_p\ 
\end{equation}
estimated at two different times: the first at $8$s of data and the second
at  $64$s.
It is evident that the quantities $B(a_p)$ are equals to zero after $1$ minute
of data.

Now to evaluate the efficiency of the estimator  LSL we verify that the
variance for the estimated coefficients  $a_p$ reaches the
Cramer-Rao (eq.~(\ref{eq:Cra})) limit.

\begin{figure}
\epsfig{file=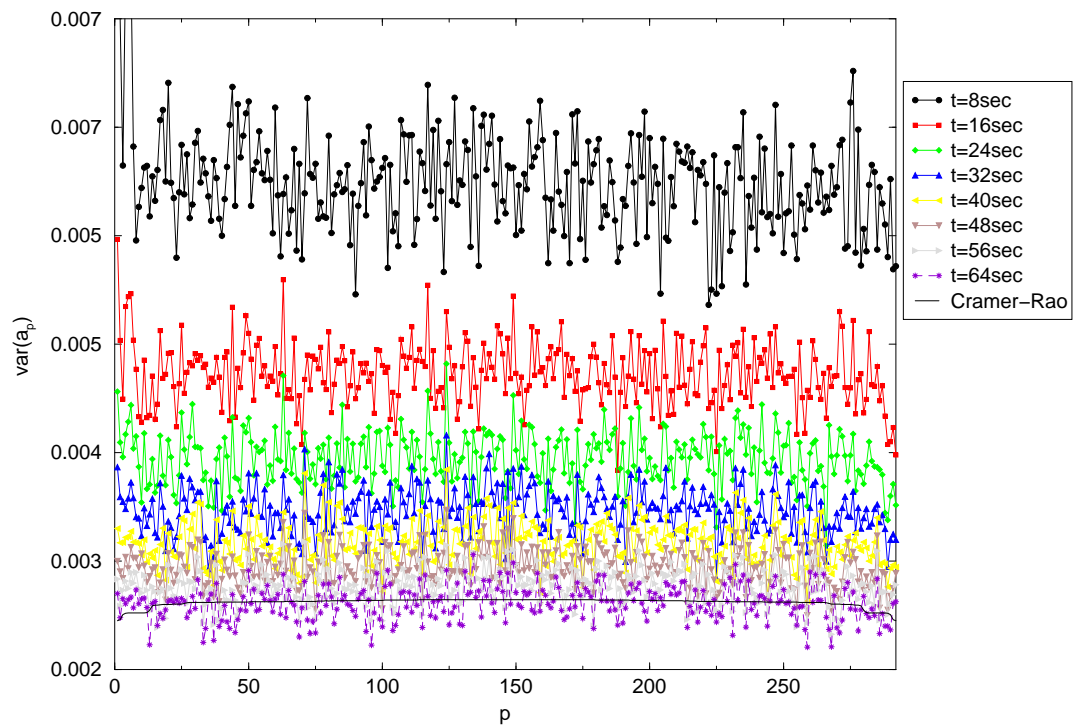,angle=270,width=0.8\textwidth}
\caption{\label{fig:CramerA}Cramer-Rao bound for LSL parameters.}
\end{figure}

In figure~\ref{fig:CramerA} we report the estimated variance for the
coefficients $a_p$ at the output of  LSL filter and the theoretical Cramer-Rao
bound.
The variance has been estimated at steps of time growing until the limit
of  $64$s. It is evident that its values become smaller and smaller
increasing the number of iteration and that it reaches the Cramer-Rao
theoretical limit~\cite{cuoco2}.

\subsection{If the noise is not a AutoRegressive process}
Suppose that our process is not an autoregressive one: does the LSL work well
in this situation? 
To verify this we simulated, as in the Durbin case, the data as an ARMA
process and test the LSL on this sequence of data.
The optimal order for the AR fit to these data  now is $338$, so we use this
order for the LSL final stage. 

\begin{figure}
\epsfig{file=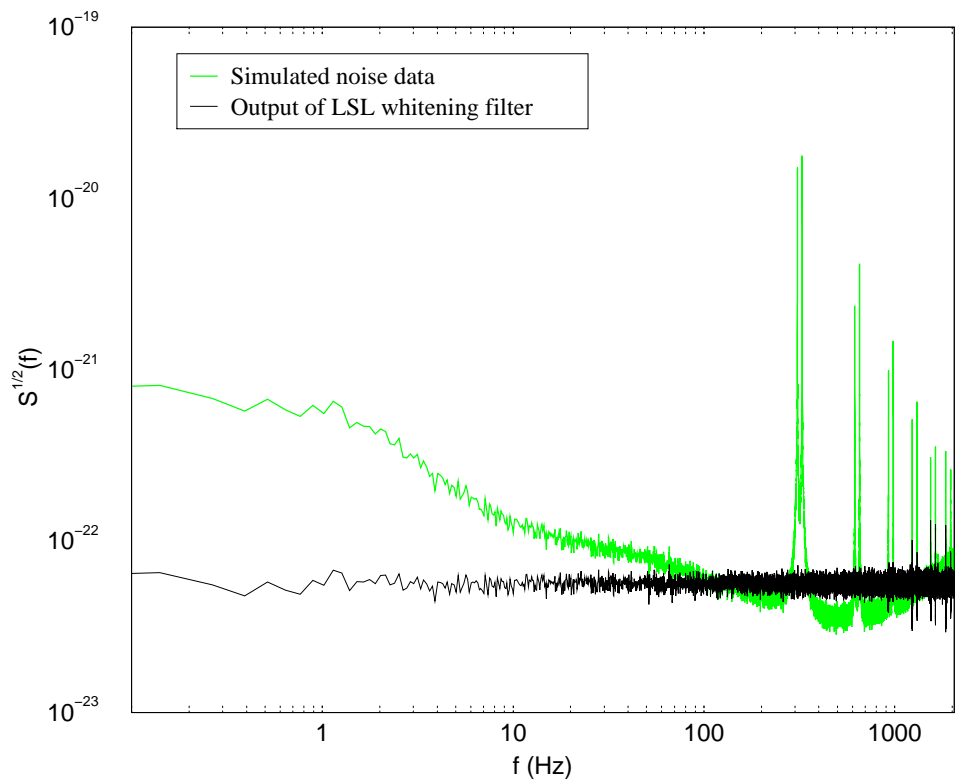,angle=270,width=0.8\textwidth}
\caption{\label{fig:lslwhite}Exit of LSL whitening filter.}
\end{figure}

In figure~\ref{fig:lslwhite} we plotted the PSD of the simulated data and at
the output of the LSL filter averaged on $100$ realizations. 
It is evident that the LSL succeeded in whitening also the ARMA sequence, even
if we use an AR fit.
In table~\ref{tab:xi} we reported the values of flatness for the outputs of
Durbin and LSL filter. 

\begin{table}
\caption{\label{tab:xi}Flatness on averaged power spectra at the input and the
  outputs of whitening filter for VIRGO-like simulated data.}
\begin{indented}
\lineup
\item[]\begin{tabular}{@{}llll}
\br 
 Simulated noise&
Durbin&
LSL&
White process\\
\mr
0.050&
 0.983&
 0.984&
 0.989\\
\br
\end{tabular}
\end{indented}
\end{table}

The values of flatness for LSL and Durbin whitening filter are similar, even
if it is evident that  LSL whitens better than  Durbin filter. 
 We think that this is related to the fact
that the adaptive filters don't need the previous estimation of autocorrelation
function; in this way if we make a mistake in estimating the autocorrelation
it will not propagate in the estimation of coefficients.

\section{Conclusion}

In this paper we addressed the problem of on-line identification of the
parameters which fitted the PSD at the output of an interferometric detector
like the VIRGO one. Moreover we face the problem of whitening on-line the
sequence  of data.

In this work we reviewed the Durbin and LSL whitening algorithms and we reported
the results of whitening on VIRGO-like noise simulated data, showing that it is
possible to obtain a whitened PSD. 
We verified that the LSL adaptive algorithm has a better performance with respect
to the static algorithm so it could be useful if we have to face with non stationary data.
 It is important to note that in selecting the order of whitening filters it
is crucial a good knowledge of the level of whiteness needed for the different
signal detection algorithms. In fact we showed that the value of flatness tends
to reach a plateau with respect to the number of parameters used in the whitening
filters while the computational cost increases proportionally to the order of
filter.
 
The procedure of whitening we described is a linear procedure that does not
destroy any part of the data. It is a reversible process that can be updated
to the level of whiteness we need. It is worth to note that the Durbin
algorithm is a 'static' procedure: we suppose to know  that we are
analyzing  only noise and we fit the parameters to perform the
whitening on the next sequence of data. In this way if in the next sequence of
data there is a signal, we don't white the signal even if we can modify its
waveform. Note that we have under control the kind of changes we made to the
signal, because we know the parameters of our whitening filter.
Instead when we use the adaptive algorithms we could in principle let the
algorithm learn also the signal buried in the noise, and whiten all the
information about it, but the learning time of the algorithm we described is
such that only a periodic signal can be captured by this algorithm. In an
incoming paper we will report the tests we made on the whitening procedure applied to sequences of data containing gravitational signals~\cite{cuoco4}.

The goodness of such filters as whitening filters has been already tested on 
data taken from prototype interferometer with very encouraging results~\cite{cuoco3, cuoco5}. 
\

\end{document}